\documentclass[twocolumn]{aastex63}
\usepackage{hyperref}
\usepackage{bm}
\usepackage{multirow}
\usepackage{gensymb}
\usepackage{enumitem}
\usepackage{tabularx}
\usepackage[flushleft]{threeparttable}
\usepackage{natbib}
\usepackage{rotating}
\usepackage{amsfonts,amsmath,amssymb}
\usepackage{graphicx}
\usepackage{color}
\usepackage{xcolor}
\usepackage{url}
\usepackage{lmodern}
\usepackage{letltxmacro}

\newcommand{\sirocco}{\textsc{sirocco}}

\newcommand{\E}[1]{\left\langle#1\right\rangle}

\def\msun{{\rm M}_\odot}

\def\hep0{{\rm HEP_{0}}}

\def\<{\,\langle\langle}
\def\>{\,\rangle\rangle}

\newcommand{\CIV}{\ensuremath{\mathrm{C\,IV}}}
\newcommand{\SiIV}{\ensuremath{\mathrm{Si\,IV}}}
\newcommand{\NV}{\ensuremath{\mathrm{N\,V}}}

\newcommand{\beq}{\begin{equation}}
\newcommand{\seq}{\end{equation}}

\renewcommand{\v}[1]{\ensuremath{\mathbf{#1}}} 
\renewcommand{\d}[2]{\frac{d #1}{d #2}} 

\newcommand{\f}{\frac}  
\newcommand{\e}[1]{\exp\left( #1 \right)} 

\usepackage{letltxmacro,amsmath}
\LetLtxMacro{\originaleqref}{\eqref}
\renewcommand{\eqref}{Eq.~\originaleqref}

\defcitealias{Dyda17}{D17}
\defcitealias{Dannen19}{D19}
\defcitealias{Dannen20}{D20}
\defcitealias{Waters21}{Waters et al. 2021}
\defcitealias{Begelman83}{BMS83}
\defcitealias{CAK}{CAK}
\defcitealias{SK90}{SK90}
\defcitealias{SV93}{SV93}
\defcitealias{KWD95}{Knigge et al. 1995}
\defcitealias{Matthews25}{M25}
\defcitealias{Luketic10v1}{L10}
\defcitealias{Giustini19}{GP19}
\defcitealias{Giustini12}{GP12}

\submitjournal{ApJL}
\shorttitle{Highly Blueshifted UV Absorption}
\shortauthors{Dannen et al.}
\begin{document}
\title{Wind Acceleration as a Driver of Detached Blueshifted Absorption in Quasar Disk Winds}
\correspondingauthor{Randall Dannen}
\author[0000-0002-5160-8716]{Randall C. Dannen}
\email{randall.dannen@unlv.edu}
\affiliation{Department of Physics \& Astronomy \\
University of Nevada, Las Vegas \\
4505 S. Maryland Pkwy \\
Las Vegas, NV, 89154-4002, USA}
\affiliation{Nevada Center for Astrophysics \\
University of Nevada, Las Vegas \\
4505 S. Maryland Pkwy. \\
Las Vegas, NV 89154, USA}
\author[0000-0002-6336-5125]{Daniel Proga}
\affiliation{Department of Physics \& Astronomy \\
University of Nevada, Las Vegas \\
4505 S. Maryland Pkwy \\
Las Vegas, NV, 89154-4002, USA}
\affiliation{Nevada Center for Astrophysics \\
University of Nevada, Las Vegas \\
4505 S. Maryland Pkwy. \\
Las Vegas, NV 89154, USA}
\author[0000-0003-0677-785X]{Paola Rodr\'iguez Hidalgo}
\affiliation{Physical Sciences Division, School of STEM\\ University of Washington Bothell\\ Bothell WA, 98011, USA}
\author[0000-0001-5506-1968]{Kara Smith}
\affiliation{Department of Physics \& Astronomy \\
University of Nevada, Las Vegas \\
4505 S. Maryland Pkwy \\
Las Vegas, NV, 89154-4002, USA}
\affiliation{Nevada Center for Astrophysics \\
University of Nevada, Las Vegas \\
4505 S. Maryland Pkwy. \\
Las Vegas, NV 89154, USA}
\author{Anna Ritchie}
\affiliation{Physical Sciences Division, School of STEM\\ 
University of Washington Bothell\\ 
Bothell WA, 98011, USA}
\author{Liliana Flores}
\affiliation{Physical Sciences Division, School of STEM\\ 
University of Washington Bothell\\ 
Bothell WA, 98011, USA}
\author{Morrigan Kalet}
\affiliation{Physical Sciences Division, School of STEM\\ 
University of Washington Bothell\\ 
Bothell WA, 98011, USA}
%
\begin{abstract}
Active galactic nuclei (AGN) unification models often emphasize the viewing angle, $i$, but $i$ alone 
does not determine quasar properties. 
This is crucial for quasar outflows: 
UV absorption in Extremely High Velocity Outflow (EHVO) quasars can reach blueshifted velocities of $\sim0.2~c$. 
In disk-wind models, both $i$ and internal wind structure shape the emergent spectrum. 
We test their interplay using biconical quasar disk-wind models 
with different acceleration lengths, $R_v$, and generate synthetic spectra 
over a range of $i$. We use Monte Carlo radiative transfer 
to account for finite continuum sources, wind attenuation, scattering, reprocessing, and emission. 
Changing $R_v$ greatly alters the ionization structure, continuum shape, and absorption-line profiles. 
At intermediate viewing angles, sight lines pass through the fastest wind. 
Even there, highly detached and blueshifted \CIV\ absorption like that observed 
in EHVO quasars appears only in models with small $R_v$. 
In these models, the gas reaches high velocity before attaining the ionization and density conditions 
favorable for \CIV. 
Models with larger $R_v$ instead produce broader, 
less detached troughs, even when the terminal velocity is very high. 
Thus, highly detached and blueshifted absorption requires both a high terminal velocity 
and small $R_v$, making such features diagnostics of disk-wind acceleration and structure. 
EHVO quasars provide a clear example, but the same principle applies more broadly to highly detached 
and blueshifted absorption in quasar outflows. 
Our results support an extended disk-wind view of AGN unification: 
$i$ selects the observed wind region, while $R_v$ shapes 
the emergent spectrum and absorption morphology.
\end{abstract}

\section{Introduction} \label{sec:intro}
Classical AGN unification models attribute much of the diversity among AGN classes
to viewing angle \citep[e.g.,][]{Antonucci93,Urry95}. In this framework,
obscured and unobscured AGNs can house similar central engines but appear
distinct when viewed through an anisotropic circumnuclear structure. 
Yet orientation alone does not account for all quasar properties, which also depend
on luminosity, Eddington ratio, black hole mass, jet power, obscuration, and
accretion-flow state.

\citet{Elvis00} extended orientation-based unification to the inner regions of
unobscured quasars by proposing a funnel-shaped disk wind. In this wind-based
picture, broad absorption lines (BALs), narrow associated UV absorbers (NALs), 
X-ray ionized absorbers, broad emission-line gas, and scattering features 
are linked to different regions or viewing angles of a common outflow. 
Thus, inclination angle, $i$ determines which part of the wind is sampled. 
However, this does not imply that all quasars share the same disk-wind structure. 
As emphasized by \citet[][\citetalias{Giustini19} hereafter]{Giustini19}, 
the existence and properties of the accretion/ejection flow depend 
on the Eddington ratio, $\Gamma\equiv L/L_{\rm Edd}$, and black hole mass, 
$M_{\rm BH}$. Moreover, winds are likely to be inhomogeneous, variable, 
and partly failed \citep[see, e.g.,][]{PSK00}.

Absorption lines directly test how $i$ and wind structure shape the
observed spectrum. However, line diversity does not by itself require a clumpy,
time-dependent, or highly inhomogeneous wind.
\citet[][\citetalias{Giustini12} hereafter]{Giustini12} showed that even
smooth, steady disk winds can produce complex, multiple, and detached absorption
troughs. A smooth wind is not necessarily monotonic along a line of sight:
velocity, density, and ionization state can vary non-monotonically through the
absorbing gas. Thus, a given trough need not represent the wind as a whole or
directly trace its global properties.

Here, we extend \citetalias{Giustini12} by replacing the point-source continuum
approximation with finite-sized X-ray and accretion-disk continuum sources.
We use the radiative-transfer code
\sirocco\ \citep[][hereafter \citetalias{Matthews25}]{Matthews25}
to include wind attenuation, scattering, reprocessing, and emission.
Holding the global wind geometry, mass-loss rate, launching radii, and terminal
velocity fixed, we vary the acceleration length, $R_v$, that sets the scale over
which the wind accelerates along streamlines.

We ask whether such smooth disk winds can explain the detached, highly
blueshifted absorption observed in some quasars
\citep{Hamann97,RodriguezHidalgo11,RodriguezHidalgo20}, including extremely
high velocity outflows \citep[EHVOs;][]{RodriguezHidalgo11}, whose UV and
optical absorption lines are blueshifted by up to $\sim0.2~c$. Thus, detached,
highly blueshifted absorption can diagnose wind launching and acceleration, with
$R_v$ acting as an additional parameter in disk-wind unification alongside $i$.

In \S\ref{sec:methods}, we outline our methods. In \S\ref{sec:results}, we
present the wind structures and spectra and compare the predicted detached,
highly blueshifted \CIV\ absorption with observed EHVOs. 
We summarize our conclusions, discuss their implications, and outline future
work in \S\ref{sec:conc}.

\begin{table}[t] 
\centering
\begin{tabular}{r | r  r l}
 \hline
   \multicolumn{4}{c}{Model Parameters} \\ \hline
      Name & Description         & Value & Units \\ \hline \hline
 $M_{co}$  & central object mass & $1\times10^{9}$* & $\msun$ \\
  $R_{*}$  & central object radius & $8.86\times 10^{14}$* & cm \\
 $\dot{M}_{acc}$ & disk mass flux & 5.00* & $\msun\,{\rm yr}^{-1}$\\
  $L_{bol}$ & bolometric luminosity & $2.36\times 10^{46}$ & erg\,s$^{-1}$ \\
   ~ & ~ & 0.188 &  $L_{\rm Edd}$\\
 $L_{X}$ & X-ray luminosity & $1 \times 10^{43}$* & erg\,s$^{-1}$ \\
 $\alpha_{X}$   & X-ray power law index & $-0.9$* & \\ 
 $R_{\rm disk}$ & maximum disk radius & 6600 & $R_{\rm ISCO}$ \\ 
 $r_1$ & inner launching wind radius & 2 & $R_{\rm ISCO}$\\
 $r_2$ & outer launching wind radius & 32 & $R_{\rm ISCO}$\\  
   $\dot{M}_{w}$ & wind mass flux & 4.75 & $\,\msun\,{\rm yr}^{-1}$ \\
   $\lambda_{s}$  & $\dot{M}_{w}$ distribution power & $-2$ & \\
   $a_{w}$  & streamline parameter (eq.\ \ref{eq:streamlines}) & 0 & \\
\multirow{2}{*}{$b_{w}$} &
\multirow{2}{*}{streamline parameter (eq.\ \ref{eq:bw_eq})} &  $b_1=0.367$ &  \\ 
          ~ &  & $b_2=0.226$ & \\
\multirow{2}{*}{$c_{w}$} &
\multirow{2}{*}{streamline parameter (eq.\ \ref{eq:cw_eq})} & $c_1=3~~~~~\,$ &  \\ 
          ~ &  & $c_2=0.188$ & \\
   $d_{w}$  & streamline parameter & 0.64 & \\ 
  $\gamma$  & streamline parameter & 1 & \\ 
  $v_{0}$ & initial velocity & 6 & km\,s$^{-1}$\\
  $v_{\infty}$ & maximum velocity & 1 & $v_{\rm esc}(r_{0})$ \\
 $R_{v}$ & wind acceleration length & 10--100 & $R_{\rm ISCO}$ \\ 
 $\alpha$  & acceleration power & 0.75 & \\
 $N_{\gamma}$ & \# of photons & $6\times10^{8}$ & \\
 $N_{r}$ & \# of $r$ grid points  & 200 & \\
 $N_{z}$ & \# of $z$ grid points & 300 & \\
 $r_{\rm min}$ & minimum $r$ grid value & 1.09 & $R_{\rm ISCO}$ \\
 $r_{\rm max}$ & maximum $r$ grid value & 3270 & $R_{\rm ISCO}$ \\
 $z_{\rm min}$ & minimum $z$ grid value & 0.0013 & $R_{\rm ISCO}$ \\
 $z_{\rm max}$ & maximum $z$ grid value & 1300   & $R_{\rm ISCO}$ \\
 $q_{r}$ & $r$--grid ratio & 1.03 & \\
 $q_{z}$ & $z$--grid ratio & 1.03 & \\
\end{tabular}
\caption{Summary of the unique and important parameters that define our wind model 
and input to \sirocco. Values marked with the asterisk (`*') indicate 
that these are the same values as used in the Quasar model in \citetalias{Matthews25}.} 
\label{table:summary-of-input}
\end{table} 

\section{Methods} \label{sec:methods}
Our aim is to isolate the effect of wind acceleration on the emergent UV spectrum.
To this end, we construct a smooth, steady, axisymmetric disk-wind model in
which the global geometry, wind mass-loss rate $\dot{M}_{w}$, inner and outer
wind-launching radii $r_1$ and $r_2$, and terminal velocity $v_\infty$ are held
fixed. 
We then vary only $R_v$, which enters through the
poloidal velocity law; see eq.~\ref{eq:vel-law}.
Changing $R_v$ modifies the velocity and density structure of the wind, 
which in turn changes the computed ionization state and emergent spectrum.
\par

We compute synthetic spectra with \sirocco, a Sobolev-based Monte Carlo
ionization and radiative-transfer code for azimuthally symmetric outflows.
Photon packets are transported in 3D through a 2.5D axisymmetric wind structure.
The code iteratively solves for the plasma state using Monte Carlo estimators of
the radiation field and then constructs observer-specific spectra using the
viewpoint technique.
\par

Although \sirocco\ can treat a variety of wind streamline geometries
\citep{Mosallanezhad25, Scepi26}, it has most often been applied to models with
straight streamlines, such as those of \citet{SV93} and \citet{KWD95}. Here, we
instead import a custom wind model with equatorial, concave streamlines similar
to the ``type 2'' streamlines shown in Fig.\ 1 of
\citetalias{Giustini12}. We adopt this geometry because it resembles the
streamlines found in line-driven disk-wind simulations
\citep[e.g.,][]{Proga98,PSK00,PK04}. We describe the imported wind model in
\S\ref{sec:wind-model} and the remaining \sirocco\ setup in
\S\ref{sec:sirocco}.
\par

\begin{figure*}
    \centering
    \includegraphics[width=\textwidth]{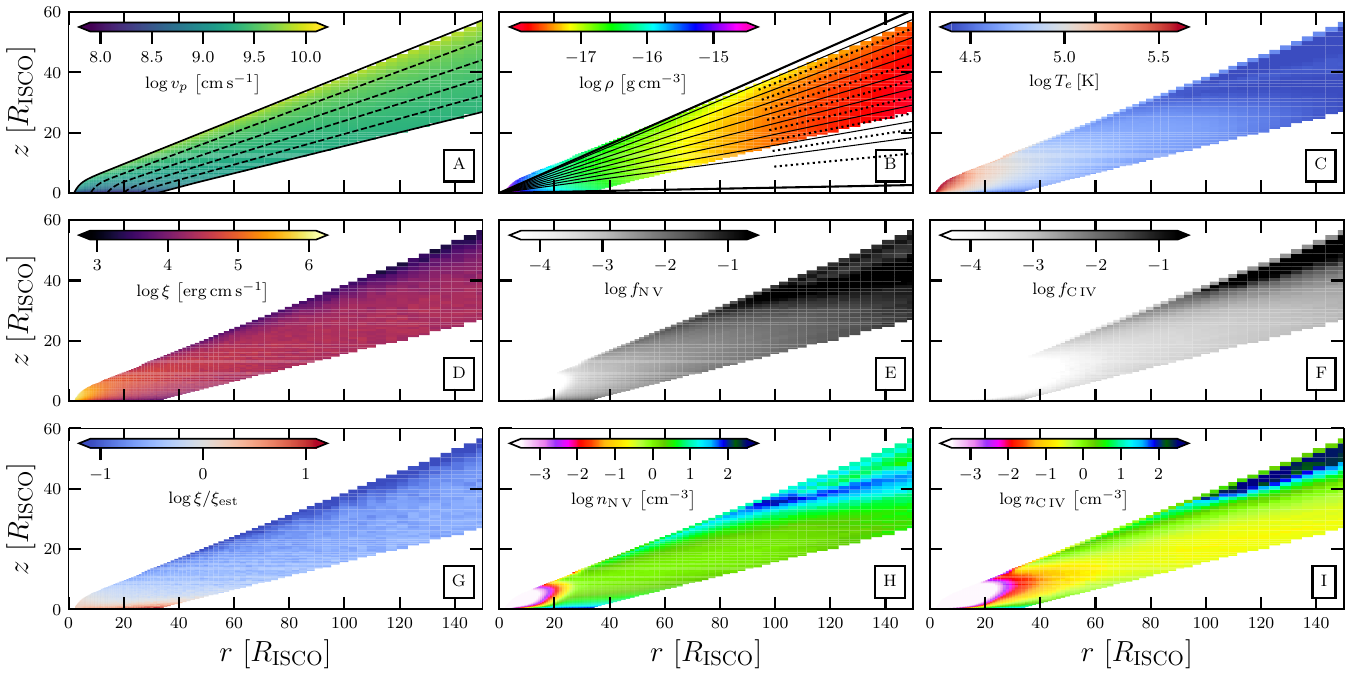}
    \caption{
    Input and output quantities from the \sirocco\ calculations for the
    fiducial model with $R_v=40~R_{\rm ISCO}$. Panel labels are shown in the
    bottom-right corner of each panel. From left to right and top to bottom, the panels show:
    A: input poloidal velocity, $v_p$, with wind streamlines overplotted
    (see eq.~\ref{eq:streamlines});
    B: input density, $\rho$, with selected lines of sight overplotted
    for $i=69^\circ$, $70^\circ$, ..., $83^\circ$, $85^\circ$, and $89^\circ$;
    C: electron temperature, $T_e$, computed by \sirocco;
    D: ionization parameter, $\xi=(4\pi)^2J_X/n$;
    E: fractional abundance of \NV, $f_\NV$;
    F: fractional abundance of \CIV, $f_\CIV$;
    G: ratio of the computed ionization parameter, $\xi$, to the optically
    thin estimate, $\xi_{\rm est}$, computed following \citet{Smith24};
    H: number density of \NV\ ions, $n_{\NV}$;
    I: number density of \CIV\ ions, $n_{\CIV}$. Units are given below the color bar in each panel.}
    \label{fig:9panel-summary}
\end{figure*}

\subsection{Wind Model}\label{sec:wind-model}
We construct an idealized, smooth, steady, axisymmetric biconical disk wind with
features motivated by line-driven disk-wind simulations: equatorial concave
streamlines and strong mass loading from small disk radii. For a streamline
launched from $(r,z)=(r_0,0)$, we set the terminal velocity to the local escape
speed,
    $v_\infty=v_{\rm esc}(r_0)=\sqrt{{2GM_{\rm co}}/{r_0}}$,
where $G$ is the gravitational constant and $M_{\rm co}=10^9\,\msun$ is the
central mass. We use $M_{\rm co}$ here to match the \sirocco\ input convention;
elsewhere, this same quantity is denoted $M_{\rm BH}$.
\par

To approximate self-shielding within a smooth steady-state model, we adopt a
modified version of the streamlines described by \citet{Luketic10v1}. The
nearly vertical inner wind shadows the outer wind, reducing overionization by
the central radiation source. Defining
$\mathcal{R}\equiv(r-r_0)/r_\mathrm{ref}$, we write
\begin{equation}\label{eq:streamlines} 
    \frac{z(\mathcal{R})}{r_\mathrm{ref}} =
    \left(a_w\mathcal{R}^2+b_w\mathcal{R}+c_w\right)
    \left[1-\exp(-d_w\mathcal{R})\right],
\end{equation}
with $r_\mathrm{ref}=R_\mathrm{ISCO}=6GM_\mathrm{co}/c^2$ 
corresponding to the ISCO radius of a nonspinning black hole.
To break self-similarity and produce a more realistic concave geometry, we let
$b_w$ and $c_w$ vary with launch radius:
\begin{equation}\label{eq:bw_eq}
    b_{w} = x^{\gamma} \left(b_2 - b_1\right) + b_1,
\end{equation} 
\begin{equation}\label{eq:cw_eq}
    c_{w} = x^{\gamma}  \left(c_2 - c_1\right) + c_1,
\end{equation}
where $x = (r_{0} - r_{1})/(r_{2} - r_{1})$. We set $\gamma=1$, so that
$b_w$ and $c_w$ vary linearly across the launching region. Choosing
$b_1>b_2$ and $c_1>c_2$ makes the streamlines increasingly equatorial with
increasing $r_0$; see the black solid and dashed lines in Panel A of
Fig.~\ref{fig:9panel-summary}. We list the full parameter set in
Table~\ref{table:summary-of-input}.
\par
To compute the poloidal velocity, $v_p$, we use the velocity law of
\citet{SV93}, also adopted in the quasar model of \citetalias{Matthews25}:
\begin{equation}\label{eq:vel-law}
    v_p(\ell) = v_0 +
    (v_\infty-v_0)
    \frac{(\ell/R_v)^\alpha}
    {1+(\ell/R_v)^\alpha},
\end{equation}
where $v_0$ is the initial poloidal velocity,
$\ell=\sqrt{z^2 + (r-r_{0})^2}$ is the poloidal distance from the launch point,
and $\alpha$ controls the shape of the acceleration law 
and can greatly effect how the gas accelerates \citep[e.g., see Fig. 4 of][]{Matthews15}.
We isolate the effect of wind acceleration by fixing
$\alpha=0.75$ and varying only $R_v$:
$R_v=10$, 15, 20, 30, 40, 80, and 100\ $R_\mathrm{ISCO}$.
This value of $\alpha$ gives rapid acceleration near the wind base, as seen in
line-driven disk-wind calculations. We initialize the wind with Keplerian
rotation at the base and conserve specific angular momentum along each
streamline, so that $v_{\phi} r = \mathrm{const}$.
\par

To set the density normalization, we adopt the same disk accretion rate as \citetalias{Matthews25},
$\dot M_\mathrm{acc}=5~\msun~\mathrm{yr}^{-1}$ 
and assume that the wind mass-loss rate is 
$\dot M_w = 0.95 \dot M_\mathrm{acc}$. 
The luminosity is $L=\eta \dot M_\mathrm{acc} c^2$, then 
adopting radiative efficiency $\eta=1/12$, we see that
our Eddington ratio is $\Gamma \equiv L/L_\mathrm{Edd}=0.188$.
We prescribe the mass flux per unit disk area as
\begin{equation}
    \dot{m}(r_0) =
    \mathcal{D} r_0^{\lambda_s}\cos\theta_0 ,
\end{equation}
where $\theta_0$ is the angle between the streamline and the disk normal at the
wind base. We choose the normalization constant $\mathcal{D}$ so that
\begin{equation}
    \dot M_w =2 
    \int_{r_1}^{r_2} \dot m(r_0)\,dA_0 ,
\end{equation}
where the factor of two includes both sides of the disk and $dA_0$ is the disk
surface-area element over the wind-launching region. We fix
$\lambda_s=-2$ for all models.
\par

We then compute the density along each streamline from mass continuity:
\begin{equation}\label{eq:rho}
    \rho(r,z) =
    \rho_0(r_0)
    \frac{\mathcal{A}(r_0,0)\,v_z(r_0,0)}
         {\mathcal{A}(r,z)\,v_z(r,z)} ,
\end{equation}
with
\begin{equation}\label{eq:rho0}
    \rho_0(r_0)=\frac{\dot m(r_0)}{v_z(r_0,0)} .
\end{equation}
Here, $A(r,z)$ is the dimensionless cross-sectional area of a stream tube:
\begin{equation}\label{eq:area}
    \mathcal{A}\left(r, z\right) =  
    \frac{1}{\sqrt{1 + \left({dr}/{dz}\right)^2}}
    \left(
    \frac{r}{r_{0}}
    \frac{\partial r}{\partial r_{0}}
    \bigg|_{z}\right) .
\end{equation}
The first factor accounts for projection along the streamline, while the second
factor accounts for streamline divergence \citep[see eq.~8 of][]{F090}.
For self-similar streamlines, such as those of \citet{Luketic10v1}, there is no
additional divergence and $\mathcal{A}\propto r/r_0$. 
In our modified streamline geometry, additional divergence gives locally
$\mathcal{A}\propto (r/r_0)^q$, with $1\lesssim q<2$, still below the
spherical-wind limit, $q=2$.
\par

We map the wind structure onto a two-dimensional spatial grid with
$N_r\times N_z$ points, spanning $r_\mathrm{min}\le r\le r_\mathrm{max}$ and
$z_\mathrm{min}\le z\le z_\mathrm{max}$. We use geometric spacing,
$d r_{i+1}=q_r d r_i$ and $d z_{i+1}=q_z d z_i$, and choose the spacing so that
the electron-scattering optical depth across a grid cell remains
$\lesssim 10$.
\par
\subsection{\sirocco\ Specifics}\label{sec:sirocco}

We import each wind model into \sirocco\ and compute the ionization balance,
thermal structure, and emergent spectra. We use the same ionizing continuum,
disk-radiation prescription, system parameters, and atomic data as in the
quasar model of \citetalias{Matthews25}. For each model, we run two full
ionization iterations, each consisting of 30 cycles. Across these cycles, we
propagate $N_{\gamma}=6\times10^{8}$ photon packets, distributed
logarithmically among the cycles.
With this setup, more than 95\% of the wind cells satisfy the \sirocco\
convergence criterion for the thermal and ionization state.

\par

We then compute the emergent spectra by propagating at least
$5\times10^{6}$ photon packets for each model. We treat line transfer with the
``escape\_prob'' approach described in the \sirocco\ documentation. We compute
spectra over 60--6000~\AA\ using 12,000 logarithmically spaced bins. Near
1200--1500~\AA, this corresponds to a typical bin spacing of
$\Delta\lambda\sim2$--3~\AA, or
$\Delta v\sim400$--600~km~s$^{-1}$. We compute spectra for 20 viewing
inclinations: $i=40^\circ$, $60^\circ$, $68^\circ$, $69^\circ$, $\ldots$,
$82^\circ$, $83^\circ$, $85^\circ$, and $89^\circ$. We measure $i$ from the
disk rotation axis, so that $i=0^\circ$ is pole-on.
\par

\begin{figure*}
    \centering
    \includegraphics[width=\textwidth]{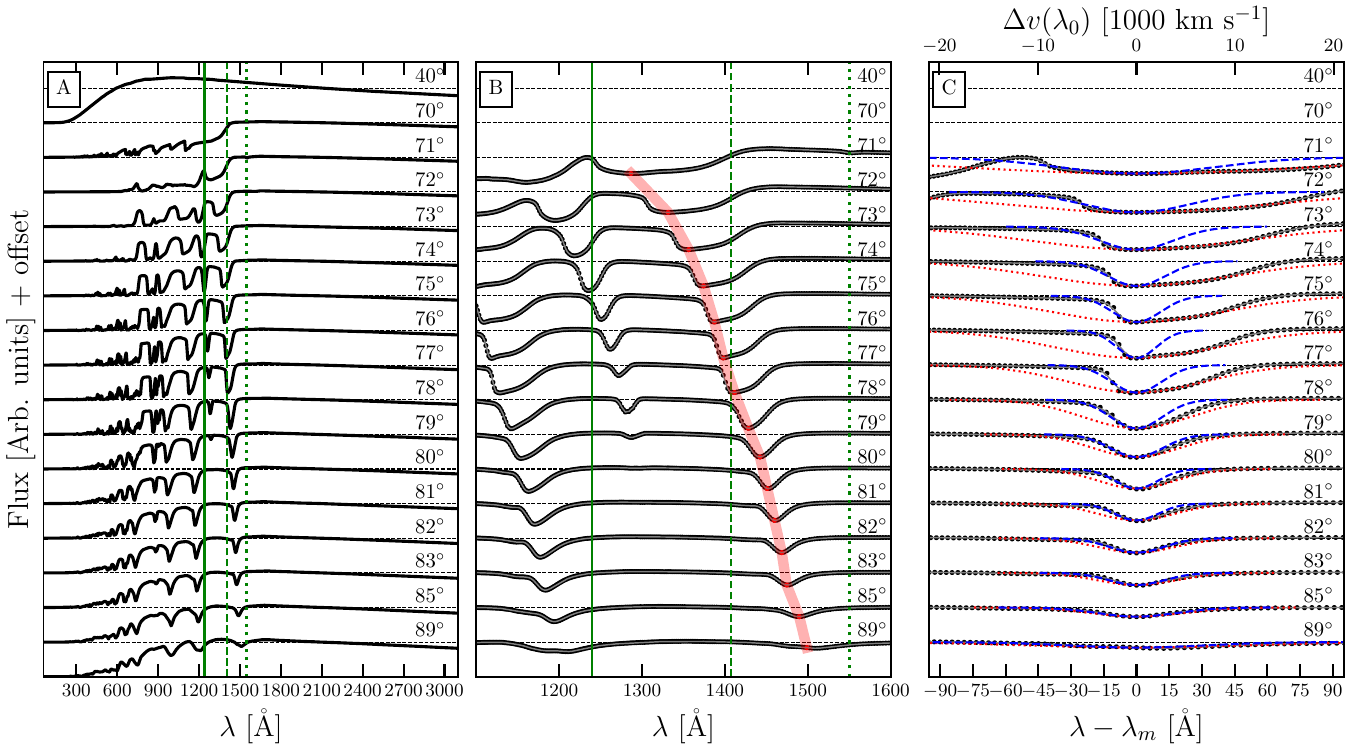}
    \caption{Synthetic spectra as a function of inclination for the fiducial $R_v=40~R_\mathrm{ISCO}$ model. 
    The spectra are normalized and vertically offset for clarity in each panel. 
    Panel labels are shown in the top-left corner of each panel. 
    From left to right, the panels show:
    {A:} Full spectra over a wide wavelength range, normalized to the flux at $\lambda=2100$~\AA. 
    The green solid, dashed and dotted vertical lines mark the rest-frame wavelengths 
    of \NV\ $\lambda1240$, \SiIV\ $\lambda1397$, and \CIV\ $\lambda1549$, respectively. The dashed line also
    approximately marks the 0.1$c$ blueshift threshold for \CIV\ absorption, commonly
    used to identify EHVOs.
    {B:} Continuum-normalized spectra over the wavelength range containing blueshifted 
    \NV, \SiIV, and \CIV\ absorption. The continuum is modeled with a quadratic fit 
    using fluxes near 1200, 1800, and 2200~\AA. The solid red curve connects $\lambda_m$, 
    the wavelength at which \CIV\ absorption depth is maximum, for each inclination. 
    {C:} Continuum-normalized \CIV\ profiles shifted relative to $\lambda_m$. 
    The red dotted and blue dashed curves show Gaussian profiles with widths set 
    by the red- and blue-side half-widths at half maximum, $\mathrm{HWHM}_{\rm r}$ 
    and $\mathrm{HWHM}_{\rm b}$, respectively. The upper axis gives the approximate
    velocity shift, computed assuming $\lambda_m=1549$~\AA.}
    \label{fig:3panel-fid-spec}
\end{figure*}

\section{Results} \label{sec:results}

\subsection{Thermal and Ionization Structure}

Figure~\ref{fig:9panel-summary} summarizes the dynamical, thermal, and
ionization structure of the fiducial model with $R_v=40~R_{\rm ISCO}$. The
figure illustrates why moderately ionized species such as \CIV\ and \NV\ can
produce detached, highly blueshifted absorption. Panels A and B show the input
poloidal velocity and density with a handful of streamlines drawn in Panel A and 
our sampled LOS indicated in Panel B. Panels C and D show the electron temperature,
$T_e$, and ionization parameter, $\xi=(4\pi)^2J_X/n$, computed by \sirocco,
where $J_X$ is the frequency-integrated mean intensity over 1--1000~Ry and $n$
is the gas number density. Panel G compares $\xi$ with the optically thin estimate, 
$\xi_{\rm est}$, computed following \citet{Smith24}. Panels E and F show the fractional
abundances of \NV\ and \CIV, $f_{\NV}$ and $f_{\CIV}$, while Panels H and I show
the corresponding number densities, $n_{\NV}$ and $n_{\CIV}$.
\par

The innermost wind is hot, with $T_e>10^5$~K, because it is directly exposed to
coronal X-rays. Its high density and concave geometry shield much of the wind at
larger radii. Because the flow also expands more slowly than the radiation field
dilutes, with local divergence parameter $q<2$, the densest and fastest parts of
the wind reach ionization parameters well below the optically thin estimate,
with $\xi/\xi_{\rm est}$ as low as $\sim 0.02$. Near the outer wind base,
however, scattering in the upper wind enhances the ionizing radiation field,
producing the high-$\xi/\xi_{\rm est}$ region near the disk in Panel G.
\par

The hottest, directly exposed gas contains little \NV\ or \CIV\
(see Panels~H and I). Instead, these ions are concentrated in the cooler,
shielded part of the fast stream, where the flow speed is already high. 
As shown in Panels~C, D, and F, \CIV\ reaches $f_{\CIV}>0.1$ for
$7000~{\rm K}\lesssim T_e\lesssim2\times10^4~{\rm K}$ and
$300\lesssim\xi\lesssim3\times10^4$, with $\xi$ in
${\rm erg~cm~s^{-1}}$. 
This high-$f_{\CIV}$ gas occupies the wedge-shaped region of high
$n_{\CIV}$ in the upper-right part of Panel~I.
Thus, the \NV\ and \CIV\ zones lie primarily in gas that 
is detached from the wind base and already moving at high velocity. 
Sightlines through this region should therefore produce an absorption line 
whose red edge is significantly blueshifted relative to the rest wavelength. 
In the next subsection, we examine how this structure maps onto the
synthetic line profiles.
\par

\begin{figure*}
    \centering
    \includegraphics[width=\textwidth]{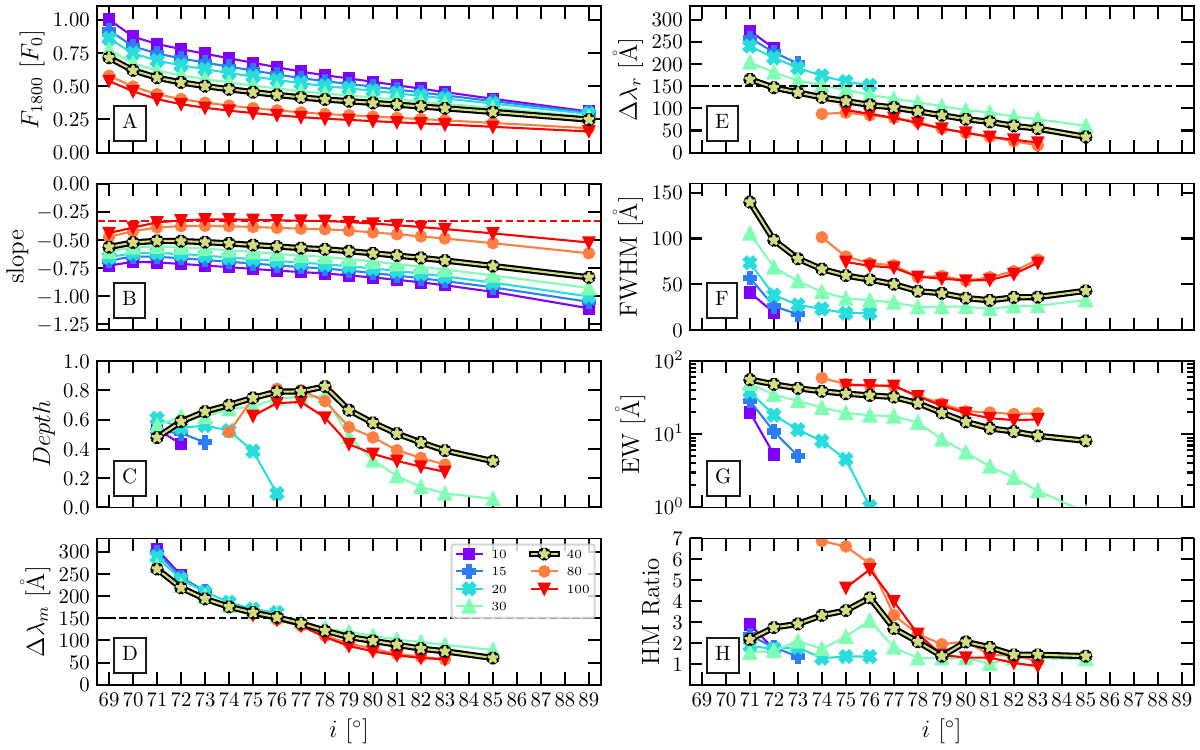}
    \caption{Properties of the continuum and \CIV\ $\lambda1549$ absorption feature as
    functions of inclination for seven wind acceleration lengths. 
    Panel labels are shown in the bottom-right corner of each panel.
    The legend in Panel D identifies the color and symbol corresponding to each value of $R_v$.
    Outlined symbols emphasize the fiducial model with $R_v=40~R_{\rm ISCO}$. 
    In Panels C--H, points are omitted when the absorption feature is absent 
    or when the line diagnostic cannot be defined unambiguously.
    From top to bottom and left to right, the panels show:
    {A:} continuum flux at $\lambda=1800$~\AA, normalized to the value for the
    $R_v=10~R_\mathrm{ISCO}$, $i=70^\circ$ model;
    {B:} continuum slope calculated using eq.~\ref{eq:powerlaw-slope-1800-2200};
    {C:} maximum absorption depth, {\it Depth};
    {D:} blueshift of the wavelength of maximum absorption depth,
    $\Delta\lambda_{m}=\lambda_{0}-\lambda_{m}$, where $\lambda_{0}$
    is the rest-frame wavelength of \CIV. 
    The black dashed line marks the EHVO threshold for \CIV;
    {E:} line detachment,
    $\Delta\lambda_{r}=\lambda_{0}-\lambda_{r}$, where $\lambda_{r}$ is the wavelength on
    the red side of the absorption feature at the point where the flux drops by half the max absorption;
    {F:} full width at half maximum, FWHM;
    {G:} equivalent width, EW;
    {H:} line asymmetry, defined as $\mathrm{HWHM}_\mathrm{r}/\mathrm{HWHM}_\mathrm{b}$.}
    \label{fig:multi-model-vert-summary}
\end{figure*}

\begin{figure*}
    \centering
    \includegraphics[width=\textwidth]{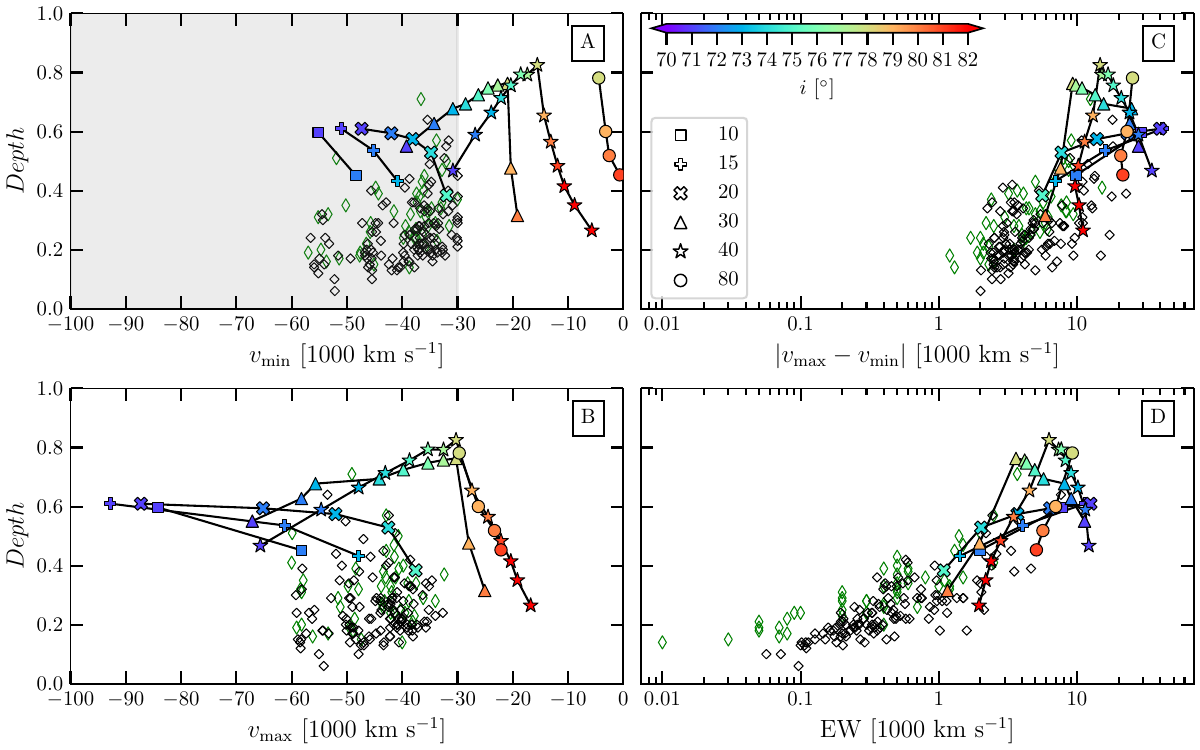}
    \caption{Comparison between model \CIV\ absorption diagnostics and observed EHVOs. 
    Filled symbols show the \sirocco\ models, with symbol shape indicating
    $R_v$ as in Fig.~\ref{fig:multi-model-vert-summary} and interior color
    indicating inclination, as shown by the color bar in Panel B. 
    The diamonds show EHVO measurements 
    from the SDSS DR9 \citet{RodriguezHidalgo20} and DR16 samples in R. Candeleria-Stoner et al. (2026, in prep).
    Panel labels are shown in the bottom-right corner of each panel.
    The panels show the maximum absorption depth, 
    {\it Depth}, as a function of:
    {A:} minimum outflow velocity, $v_\mathrm{min}$, defined by the red edge 
    of the absorption trough where the normalized flux first drops below 0.9, 
    the shaded region indicates the velocity range to be considered an EHVO;
    {B:} maximum outflow velocity, $v_\mathrm{max}$, defined by the blue edge 
    where the normalized flux recovers above 0.9;
    {C:} line width as the difference between $v_\mathrm{max}$ 
    between $v_\mathrm{min}$, and 
    {D:} equivalent width, EW. with units 
    1000 km s$^{-1}$.
    }
    \label{fig:observe-multi-scatter}
\end{figure*}
\subsection{Synthetic Spectra}
Figure~\ref{fig:3panel-fid-spec} shows synthetic spectra for the fiducial
$R_v=40~R_{\rm ISCO}$ model at selected inclinations. Panel A shows nearly the
full wavelength range, normalized to the continuum flux at
$\lambda=2100$~\AA. The green solid, dashed, and dotted vertical lines mark the
rest wavelengths of the \NV, \SiIV, and \CIV\ resonance lines at
$\lambda=1240$, 1397, and 1549~\AA, respectively. The dashed line also lies
near the wavelength of \CIV\ blueshifted by $0.1c$, and therefore marks the
approximate EHVO threshold for \CIV\ absorption (see the shaded region 
in Panel~A of Fig.~\ref{fig:observe-multi-scatter}).
\par

Panel B focuses on the blueshifted \CIV\ absorption. In Panels B and C, we
renormalize the spectra with a quadratic continuum fit anchored near
$\lambda\simeq1200$, 1800, and 2200~\AA. As $i$ decreases from $89^\circ$, the
\CIV\ absorption shifts blueward and, for some inclinations, crosses the
$0.1c$ EHVO threshold. We define the absorption depth as
${\it Depth}=1-F_{\lambda}/F_{\lambda,\mathrm{cont}}$. The red curve connects
$\lambda_m$, the wavelength of maximum depth, for each inclination. The spectra
also show blueshifted \SiIV\ $\lambda1397$ absorption for
$i=71$--$76^\circ$ and \NV\ $\lambda1240$ absorption at all inclinations. These
features shift together with \CIV\ while preserving the expected wavelength
separations, supporting the line identification.
\par

Panel C compares the \CIV\ profiles after shifting each spectrum relative to
$\lambda_m$. We characterize each profile by the distances from $\lambda_m$ to
the half-maximum depth on the red and blue sides, $\mathrm{HWHM}_\mathrm{r}$ and
$\mathrm{HWHM}_\mathrm{b}$, so that
$\mathrm{FWHM}=\mathrm{HWHM}_\mathrm{r}+\mathrm{HWHM}_\mathrm{b}$. For comparison, we plot two Gaussian
profiles for each synthetic line. Both have the same $\lambda_m$ and maximum
depth as the corresponding \sirocco\ profile, but their widths are set
separately by $\mathrm{HWHM}_\mathrm{r}$ and $\mathrm{HWHM}_\mathrm{b}$.
\par

The profiles for $i\geq82^\circ$ are nearly symmetric, with
$\mathrm{HWHM}_\mathrm{r}/\mathrm{HWHM}_\mathrm{b}\simeq1$, and close to Gaussian. 
Their widths greatly exceed the thermal widths and do not result 
from imposed turbulence or other non-thermal broadening. 
Instead, they come from the range of projected wind velocities along the line of sight. 
At smaller inclinations, the profiles
become asymmetric: the blue wings remain approximately Gaussian over a broader
inclination range, while the red wings deviate more strongly from a Gaussian
shape.
\par


\subsection{Dependence on Acceleration Length}
Figure~\ref{fig:multi-model-vert-summary} summarizes seven wind models that
differ only in acceleration length, $R_v$. Changing $R_v$ affects both the UV
continuum and the absorption line profiles.

Panels A and B show the continuum flux and slope as functions of inclination 
and $R_v$. Panel A shows the 1800~\AA\ continuum flux, 
normalized to the maximum value in the model grid, 
which occurs for $R_v=10~R_\mathrm{ISCO}$ and
$i=70^\circ$. 
Panel B shows the continuum slope,
\begin{equation}\label{eq:powerlaw-slope-1800-2200}
\mathrm{slope} =
\frac{\log\left(F_{2200}/F_{1800}\right)}
{\log\left(2200\mathrm{\AA}/1800\mathrm{\AA}\right)},
\end{equation}
where $F_{1800}$ and $F_{2200}$ are the continuum fluxes at $\lambda=$ 1800 and
2200~\AA. The continuum flux decreases with inclination more slowly than the
flat-disk expectation, $F\propto\cos i$, because wind scattering redirects
continuum photons into high-inclination sight lines \citep[see also][]{Sim10}.
The same scattering produces $\xi/\xi_{\rm est}>1$ near the wind base. 
At a fixed mass-loss rate, increasing $R_v$ makes the wind accelerate more gradually,
raising the density and optical depth over much of the flow. As a result,
$F_{1800}$ decreases at fixed inclination by about a factor of two from
$R_v=10~R_{\rm ISCO}$ to $R_v=100~R_\mathrm{ISCO}$. The UV continuum also steepens
with increasing $R_v$ and~$i$.
\par

Panels C--H show diagnostics of the \CIV\ absorption feature. Missing points
mark spectra in which the feature is absent or cannot be defined unambiguously.
The diagnostics are maximum absorption depth, {\it Depth} (Panel C); blueshift
of maximum depth, $\Delta\lambda_{m}=\lambda_{0}-\lambda_{m}$ (Panel D); line
detachment, $\Delta\lambda_r=\lambda_0-\lambda_r$ (Panel E); FWHM (Panel F);
equivalent width (Panel G); and profile asymmetry,
$\mathrm{HWHM}_\mathrm{r}/\mathrm{HWHM}_\mathrm{b}$ (Panel H). Here $\lambda_{0}=1549$~\AA\ is the
\CIV\ rest wavelength, and $\lambda_{r}$ is the red-side wavelength where the
flux first falls below 90\% of the continuum.
\par

When \CIV\ absorption is present, decreasing $i$ shifts $\lambda_m$ blueward and
increases $\Delta\lambda_m$.
This shift depends
only weakly on $R_v$. Other line properties depend more strongly on $R_v$ and
can vary nonmonotonically with inclination. For example, {\it Depth} peaks near
$i\sim78^\circ$ for $30\leq R_v/R_\mathrm{ISCO}\leq100$, whereas the smaller-$R_v$
models show \CIV\ absorption only when the sight line crosses the densest part
of the fast stream, near $i\sim72^\circ$.
\par
The most rapidly accelerating models produce the most detached absorption, with
$\Delta\lambda_{r}$ reaching $\sim250$~\AA. These highly detached features are also relatively narrow and weak. All models share the same terminal velocity, so high terminal velocity alone is not sufficient to explain the detachment: we must also consider the acceleration history. The wind must accelerate rapidly enough for the appropriate ionization and density conditions to occur in fast-moving gas.
\par
\subsection{Comparison with Observations}

Figure~\ref{fig:observe-multi-scatter} compares our \sirocco\ models
(filled symbols) with EHVO measurements from SDSS DR9
\citep[green diamonds;][]{RodriguezHidalgo20} and DR16
(black diamonds; R. Candeleria-Stoner et al., 2026 in prep.). 
Each panel shows the maximum
absorption depth, \textit{Depth}, as a function of a diagnostic measured
consistently for the models and observations: the red- and blue-side velocities
at which the normalized flux crosses 0.9,
$v_{\rm min}$ and $v_{\rm max}$; the corresponding velocity width,
$\vert v_{\rm min}$ and $v_{\rm max}\vert$; and the equivalent width, EW, 
similar to how these quantities are computed in \citet{RodriguezHidalgo20}. 
Note that in Figure~\ref{fig:observe-multi-scatter} we omit data from our $R_{v}=100$. 
This is due to the more restrictive requirements to be included, we require that the absorption recover above 0.9 using our normalization method to be properly compared with the SSDS DR9 and DR16 EHVO measurements. This is different from the requirement to be shown in Figure~\ref{fig:multi-model-vert-summary}, where we only needed to recover to the half maximum.
\par

\par
By definition, EHVOs have outflow velocities exceed $30{,}000~\mathrm{km~s^{-1}}$, 
corresponding to a \CIV\ wavelength shift 
of $\sim 150$~\AA. Models with $R_v\leq30~R_\mathrm{ISCO}$ reach this
high-velocity regime and overlap with the deepest, largest-EW EHVO features,
consistent with the strongly absorbing end of the population. 
Models with $R_v\leq20~R_\mathrm{ISCO}$ even reach maximum outflow velocities 
of $\sim$0.3$c$ (see Fig.~\ref{fig:observe-multi-scatter} Panel B), 
which have been observed in \CIV\ absorption in two cases \citep{Hamann18,Seaton26}.
However, the observed sample tends to include features that are weaker, shallower, 
and narrower than simulated results, with
${\it Depth}<0.4$ and $\mathrm{EW}<7000~\mathrm{km~s^{-1}}$, beyond the range
covered by our current model grid.
Our \sirocco\ models best replicates the detected EHVOs that have broadest and deepest absorption lines.

\par

The main wind feature that causes relatively narrow, highly blueshifted absorption 
is a thin zone with a comparably high $n_{\CIV}$ in the outflow, 
which has already reached much of its terminal velocity, similar 
to the wedge-shaped high $n_{\CIV}$ region shown 
in Panel~I of Fig.~\ref{fig:9panel-summary}. 
The fiducial $R_v=40~R_\mathrm{ISCO}$ model produces less blueshifted and
broader absorption than the $R_v\leq30~R_\mathrm{ISCO}$ models. Its
\CIV-bearing gas extends closer to the disk and spans a wider range of projected
velocities.
We chose the $R_v=40~R_\mathrm{ISCO}$ model as the fiducial case because it
produces \CIV\ absorption over the widest range of inclinations, not because it
produces the narrowest or most detached lines. 
\par
\section{Discussion and Conclusions} \label{sec:conc}

We have computed the photoionization structure and emergent spectra of a smooth,
biconical disk wind launched from an accretion disk around an SMBH. Holding the
global wind geometry, mass-loss rate, launching radii, and terminal velocity
fixed, we varied only the acceleration length, $R_v$, to isolate its effect on
the UV continuum and absorption-line profiles.

Using the Monte Carlo radiative-transfer code \sirocco, we extend the
systematic disk-wind study of \citetalias{Giustini12} beyond the point-source
approximation by including finite continuum sources, attenuation, scattering,
reprocessing, and emission. We find that even smooth winds can produce diverse
spectra because shielding and scattering create strongly direction-dependent
thermal and ionization structures.

The acceleration length has a major effect on the observable spectra. Slowly
accelerating winds keep more material dense and moderately ionized over a broad
velocity range, producing broader and less detached absorption. Rapidly
accelerating winds can instead confine \CIV-bearing gas to a compact,
fast-moving zone, producing relatively narrow, detached, highly blueshifted
absorption. Therefore, our comparison with EHVO quasars favors winds 
that reach a large fraction of their terminal velocity over a short distance.

These results support an extended disk-wind view of AGN unification in which
viewing angle is important but insufficient to determine the observed spectrum.
Our synthetic spectra show that inclination, $i$, determines which part of the
wind is viewed, while $R_v$ shapes the ionization structure, continuum shape,
and absorption-line morphology along that line of sight. Detached, highly
blueshifted absorption may therefore diagnose wind launching and acceleration.

We prescribe the streamline geometry and velocity field rather than calculate the
wind dynamics. Our results, therefore, constrain the values of $R_v$ required
within the adopted geometry but do not uniquely identify the driving mechanism.
The concave geometry has the important advantage that the nearly vertical inner
wind shields the outer flow from ionizing radiation, allowing moderately ionized
gas to persist at high velocity. 
Such concave, equatorial streamlines, together with the small $R_v$ favored by
the strongest EHVO features, resemble radiation- or thermally driven
hydrodynamic winds more than classic self-similar magnetocentrifugal winds. 
The latter generally produce more polar, convex streamlines that collimate with
distance \citep[e.g.,][]{BlandfordPayne,Everett05}.
The diagnostic power of detached, blueshifted absorption as a probe of $R_v$ 
depends on the self-shielding enabled by our concave streamline geometry. 
Adopting more polar or convex geometries, the same relationship between $R_v$ 
and detachment may not hold, because the shielding material will be exposed 
to more of the corona and change the distribution of the \CIV\ zones, 
maybe destroying them entirely.
More generally, magnetic fields can
produce diverse wind properties: magnetic-pressure-dominated flows may be denser
and slower \citep[e.g.,][]{Proga03,Waters18}, whereas magnetocentrifugal
launching from sufficiently small radii can yield rapid acceleration and high
velocities \citep[e.g.,][]{Fukumura22}.

Radiation magnetohydrodynamic (MHD) and general relativistic MHD (GRMHD) 
simulations already generate outflows through the coupled action of radiation,
magnetic fields, and the accretion flow \citep[e.g.,][]{Jiang24,Zhang25}.
Computing synthetic UV spectra from such models, 
using realistic quasar parameters with a consistent ionization and
radiative-transfer treatment, would enable direct comparisons among
radiation-driven, magnetically driven, and hybrid winds. Differences in
acceleration, density stratification, ionic structure, and profile asymmetry
could then test whether detached absorption distinguishes among the underlying
launching mechanisms.

We varied only $R_v$ and therefore did not explore the full range of quasar-wind
properties, 
only replicating the strongest and widest absorption profiles. 
Varying the mass-loss rate, ionizing continuum, and wind geometry
can drastically alter the intrinsic radiation field and computed ionization structure and absorption line profiles and 
may extend our models toward the shallower, lower-EW part of the EHVO
distribution. Future work should also examine clumping and time dependence and
compare predictions from dynamical wind models with EHVOs, BALs, mini-BALs,
NALs, and X-ray-detected ultrafast outflows.

\section{Acknowledgments}
We thank Joshua Key, Sergei Dyda, Margherita Gustini, and Tim Waters for useful discussions. 
We thank James Matthews, Knox Long, and Christian Knigge for their early
support and development of
\sirocco\footnote{\url{https://github.com/sirocco-rt/sirocco}} \citepalias{Matthews25}.
R.D. and D.P. acknowledge support from NSF grant AST-2107883 
and from NASA grant HST-GO-16196. 
P.R.H. and L.F. acknowledge support from NSF grant AST-2107960, SDSS FAST IV program, 
and the UW Mary Gates Research Scholarship.
R.D. acknowledges the use of and greatly appreciates the publicly available software packages 
\textsc{scipy}\footnote{\url{https://scipy.org/}} \citep{2020scipy} and
\textsc{matplotlib}\footnote{\url{https://matplotlib.org/}} \citep{Hunter:2007}.
\bibliography{progalab-shared}
\end{document}